# Einstein's Equation in Pictures

## Matthew Frank

March 15, 2002

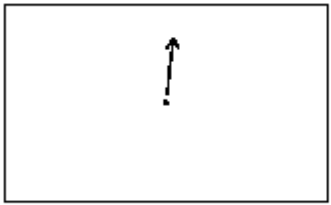
Start with a unit timelike vector v at a point p.

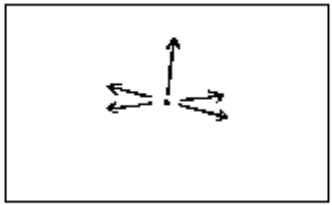
Take all the spacelike vectors orthogonal to v.

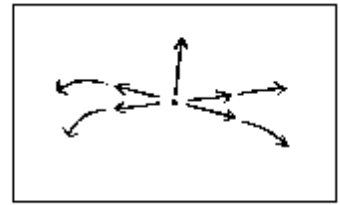
Extend these spacelike vectors into geodesics.

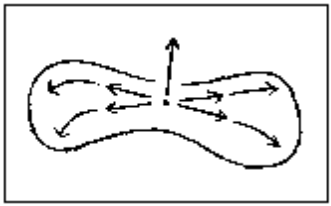
These geodesics form a 3-d hypersurface.

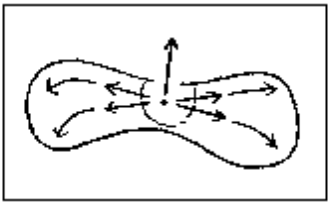
Take small balls in this hypersurface, the points within distance r of p.

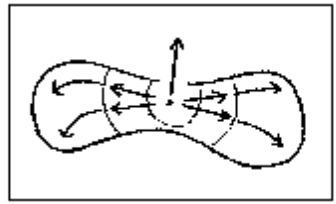
Consider the volume V(r) of these balls as a function of r.

Einstein's equation says that energy is the curvature of space. What does this mean? In terms of the above pictures, it can be expressed as:

$$16\pi \text{ (energy density as measured by } v) = \lim_{r \to 0} \frac{15}{r^2}(1 - \frac{V(r)}{\frac{4}{3}\pi r^3})$$

for each unit timelike vector v. (And that's it!)

## How To Use This Paper

I hope that both people new to and people familiar with general relativity will read this paper. The section on preliminaries is intended primarily for those new to general relativity; I hope that physical intuition will carry people most of the way through that section, but mathematicians may find it useful to know that the four-dimensional space-time metric has signature -+++. The section on comparisons with other formulations and the appendices are intended primarily for those already familiar with general relativity; I hope that these people will appreciate the novelties of this approach.

# Preliminaries

The background to this is that general relativity treats space-time as a four-dimensional manifold with a metric and an energy tensor. Since space-times are four-dimensional, space-time diagrams conventionally omit one dimension.

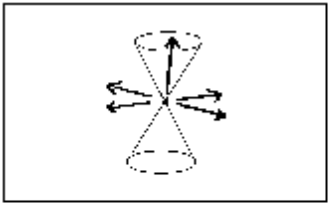
Timelike vectors are the ones that go inside a light cone; spacelike vectors, outside.

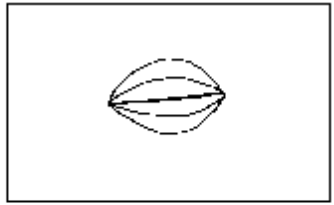
A geodesic is a path whose length is insensitive to small perturbations.

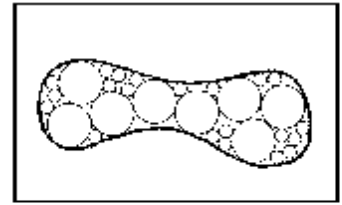
Volumes can be estimated by filling them with small, nearly Euclidean spheres.

It is the metric which makes the pictures and the right hand side of Einstein's equation meaningful. The metric determines (among other things) the possible paths of light rays, and these form a light cone. The vectors which point outside the light cone are called spacelike; the vectors on the light cone are called null. The vectors which point inside the light cone are called timelike; they represent the possible directions of inertial observers, and are often identified with observers. (This is treated in more detail in chapter 6 of Taylor and Wheeler's Spacetime Physics.)

The metric also determines distances and lengths of curves. A geodesic is a locally straightest path, a path $\gamma_0$ such that for any continuous variation of curves $\gamma_u$, the derivative (d/du)(length of $\gamma_u$) vanishes at u=0. Note that any vector can be followed into a geodesic path, as indicated in the third picture for Einstein's equation.

With the notion of distance comes a notion of volume: there is a unique volume function such that, for all sufficiently small r and all points p, $(4/3)\pi(r^3-r^4) < V(B(p,r)) < (4/3)\pi(r^3+r^4)$, where B(p,r) is the ball of radius r around p. (This is proved in the appendix.) In fact these volumes do not deviate from Euclidean volumes at the $r^4$ level at all, and the $r^5$ deviation from Euclidean volume (as on the right hand side of Einstein's equation) measures the curvature of the manifold at the point p. However, these $r^4$ inequalities are strict enough to give a procedure for estimating volumes to within an arbitrarily small factor of 1±ε: simply divide the region into countably many balls of radius at most ε and sum their Euclidean volumes.

It is the energy tensor which makes the left-hand side of the equation meaningful. This concept of "energy density in the direction of v" or "local energy as measured by the observer v" is best illustrated by examples. In a vacuum, the energy in any direction is always 0. An electromagnetic field has total energy (E · E - B · B)/8π where E and B are the vectors for the electric and magnetic fields as measured by the observer v. A perfect fluid is observed by v to have energy $(\rho + P)(u \cdot v)^2 - P$, where $\rho$ is the density of the fluid, P is its pressure, and u is its direction.

Having this at hand makes it possible to go through an example of Einstein's equation in detail; this may in particular help clarify the dimensionality of the vector spaces and manifolds. Consider Minkowsi space (the space-time of special relativity), coordinatized as (x,y,z,t). Also consider an observer at the origin moving in the t-direction. For this observer, the spacelike vectors are all the vectors which have no t-component. The geodesics form the 3-d hypersurface t=0. The ball of radius r is all points (x,y,z,0) with $x^2 + y^2 + z^2 < r^2$, and it has volume exactly $(4/3)\pi r^3$. Hence the curvature of the hypersurface is 0, as it should be since Minkowski space is a vacuum.

## Comparisons with the usual statement of Einstein's equation

The statement of Einstein's equation here makes precise "energy is the curvature of space", while the usual statement makes precise "energy-momentum is the curvature of space-time". The right hand side of the statement here gives the scalar curvature of the indicated hypersurface, and that hypersurface is the natural "space" associated to the given observer.

It is possible to do calculations that are guided by these pictures rather than by the standard Christoffel symbols or differential forms. The key is to coordinatize several things and represent them as power series in the distance from the initial point: first the geodesics from that point, then the metric of the resulting hypersurface, then the volumes of the geodesic balls in it. The curvature is proportional to the $r^5$ term in the power series for the volume. Even if the metric is only $C^3$ and not analytic, there are enough meaningful terms in these power series to allow this calculation of the curvature. I have used these techniques to rederive the Schwarzschild and Robertson-Walker solutions in this format. Unfortunately, the calculations by this method require calculations much longer than the usual ones, even when all of them are automated in Mathematica.

This statement of Einstein's equation is equivalent to the usual one. (Indeed, it is very close to the statement of Einstein's equation given by Misner, Thorne, and Wheeler on p. 515.) Unfortunately, while the usual statement is clean, and these pictures are clean, proving the equivalence of the two is somewhat messy. The following is a sketch of a proof in three steps, using geometrized units c=G=1.

- First, the usual statement $8\pi T_{ab} = G_{ab}$ is equivalent to the claim that $8\pi T_{ab} v^a v^b = G_{ab} v^a v^b$ for all unit timelike vectors v (where $T_{ab} v^a v^b$ is what is referred to above as the energy density in the inertial frame of v). This equivalence is a matter of linear algebra, using the facts that T and G are symmetric tensors and that the unit timelike vectors from a spanning set for the space of all tangent vectors.
- Second, $G_{ab} v^a v^b$ is half the scalar curvature of the indicated hypersurface; this is a special case of the Gauss-Codazzi equations without the terms for extrinsic curvature, because the extrinsic curvature of the hypersurface vanishes at p. These equations are discussed in Wald, sec. 10.2, and the appendix proves the vanishing of the extrinsic curvature can also be proved using the machinery of that section.
- Third, the scalar curvature of the hypersurface is given by the limit of $(15/r^2)(1-V(r)/V_{Euc}(r))$ as r goes to 0; for this, see Cartan, sec. 234.

## Conclusion

Two advantages of this presentation of Einstein's equation may be obvious:

- It is very pictorial.
- It requires much less of the standard mathematical apparatus: no curvature tensors (almost no tensors at all), and no parallel transport / derivative operators / affine connections.

Let me also call attention to a few advantages which may not be obvious.

- It may be easier to appraise the standard presentation of Einstein's equation given another presentation as different as this one; the standard presentation may seem less geometrically compelling but computationally not so bad by comparison. (For another alternative presentation, see Baez.)
- This presentation brings the geometry of general relativity closer to the ideal of a synthetic differential geometry set out by Herbert Busemann. (That was some of my inspiration for this project.)
- Most optimistically, this presentation of Einstein's equation (or slight variants) may be meaningful in physical theories which do not treat space-time as a 4-dimensional Lorentzian manifold.

In any case, I will be happy if this helps people to understand Einstein's equation, or gives pleasure to those who already do.

## Bibliography and Acknowledgements

- John Baez (2001), The Meaning of Einstein's Equation, gr-qc/0103044
- Herbert Busemann (1955), Geometry of Geodesics.
- Elie Cartan (1928), Lecons sur la Geometrie des Espaces de Riemann; translated by James Glazebrook (1985) as Geometry of Riemannian Spaces.
- Charles Misner, Kip Thorne, and John Wheeler (1973), Gravitation.
- Edwin Taylor and John Wheeler (1992), Spacetime Physics.
- Robert Wald (1984), General Relativity.
- my website, http://zaphod.uchicago.edu/~mfrank/GR/einpics.html, for further appendices with the Mathematica code refered to above.

Thanks to: several people at the University of Chicago and at the 1999 Notre Dame conference on history and foundations of general relativity, an October 1999 audience at the Chicago-area philosophy of physics group, John Beem, and Stephen Wolfram.

# Appendix on Riemannian volumes

Let B(p,r) be the ball of radius r around p. Then the precise claim is:

For any compact Riemannian manifold, the usual volume is the unique countably additive measure such that for all sufficiently small r and all points p,
$(4/3)\pi(r^3-r^4) < V(B(p,r)) < (4/3)\pi(r^3-r^4)$.

Proof of existence: By the result of Cartan, the limit as r approaches 0 of $(15/r^2)(1 - V(B(p,r))/(4/3)\pi r^3)$ is equal to the curvature of the manifold at p; in particular it exists and is finite. Hence $f(p,r)=[V(B(p,r)) - (4/3)\pi r^3] / [(4/3)\pi r^4]$ is a continuous function of p and r which is 0 when r is 0. Since the manifold is compact, there is some $\delta$ such that for all $r<\delta$, $|f(p,r)|<1$; this yields the claimed inequality.

Proof of uniqueness: Say V and V' are both volume functions satisfying the above inequalities. Then, for any S and any $\varepsilon$, we can cover S by countably many balls of radius at most $\min(\delta,\varepsilon)$ with non-overlapping interiors. The boundaries of these balls will be of measure 0 according to both V and V'. Hence V(S) is the sum of V of the balls, and likewise for V'. For each ball B, V(B) and V'(B) both differ from Euclidean volume by within a factor of $1\pm\varepsilon$, so V(B) and V'(B) differ from each other by within a factor of $(1\pm\varepsilon)^2$. Hence V(S) and V'(S) differ from each other by within a factor of $(1\pm\varepsilon)^2$ for each $\varepsilon$, and so V(S)=V'(S). QED.

# Appendix on the curvature of the hypersurface

This uses abstract index notation and several results from Wald. Let *v* be the original timelike vector at *P*, and let **H** be the corresponding spacelike hypersurface. Let $n^a$ be a vector field (including *v*) of unit normals to **H**, and extend it beyond **H** in such a way that its integral curves are unit geodesics; this is useful in defining the extrinsic curvature $K_{ab}$.

Wald defines $K_{ab}$ as $\nabla_a n_b$. At P, $K_{ab}=0$ since its contraction with any bivector $y^a z^b$ is 0. Proof: $n^a \nabla_a n_b = 0$ because the integral curves of $n^a$ are geodesics; $n^b \nabla_a n_b = \nabla_a(n^b n_b)/2 = 0$ because $n^a$ is of unit length. So it suffices to consider spatial vectors $y^a$ and $z^b$. For any vector $w^a$ orthogonal to $n^a$, there is a geodesic vector field including $w^a$ tangent to **H**; and for any two such vector fields, their Lie bracket is also tangent to **H**. Hence $[y^b, z^b] n_b = 0$ ; using the orthogonality of $y^b$ and $z^b$ with $n_b$, this may be rewritten as $y^{[a} z^{b]} \nabla_a n_b = 0$. So it suffices to consider symmetric bivectors $y^{(a} z^{b)}$, which may in turn be reduced to those of the form $w^a w^b$. For these also, $w^a w^b \nabla_a n_b = w^a \nabla_a (w^b n_b) = 0$, where the first equality is because $w^a$ is geodesic and the second because $w^b n_b = 0$. QED.

Now $h_{ab} = g_{ab} + n_a n_b$ is the metric on **H**, and $h^a{}_b h^{bc} = h^{ac}$.
At P, Wald 10.2.23 may be written without the terms for K as:
$^{(3)}R_{abcd} = h_a{}^f h_b{}^g h_c{}^k h_d{}^j R_{fgkj}$. Contracting both sides with $h^{ac} h^{bd}$ we get
$^{(3)}R = h^{fk} h^{gj} R_{fgkj}$, which Wald 10.2.29 shows equal to $2 G_{ac} n^a n^c$.
In other words we have $^{(3)}R/2 = G_{ac} n^a n^c$, as claimed in the text of the paper.